\documentclass[aps,prb,twocolumn,preprintnumbers,final]{revtex4}

\usepackage{graphicx,amsmath,amssymb,mathrsfs,array,longtable,delarray,verbatim,epsfig}

\begin{document}

\title{Row coupling in an interacting quasi-one-dimensional quantum wire investigated using transport measurements}

\author{L. W. Smith}
\author{W. K. Hew}
\author{K. J. Thomas$^*$}
\author{M. Pepper$^\dagger$}
\author{I. Farrer}
\author{D. Anderson}
\author{G. A. C. Jones}
\author{D. A. Ritchie}

\affiliation{Cavendish Laboratory, J. J. Thomson Avenue, Cambridge, CB3 OHE,
United Kingdom}


\begin{abstract}

We study electron transport in quasi-one-dimensional wires at relatively weak electrostatic confinements, where the Coulomb interaction distorts the ground state, leading to the bifurcation of the electronic system into two rows. Evidence of finite coupling between the rows, resulting in bonding and antibonding states, is observed. At high dc source-drain bias, a structure is observed at $0.5(2e^2/h)$ due to parallel double-row transport, along with a structure at $0.25(2e^2/h)$, providing further evidence of coupling between the two rows.\\

PACS number(s): 73.21.Hb, 72.20.-i, 73.23.Ad

\end{abstract}

\maketitle

Electrostatic confinement of a two-dimensional electron gas (2DEG) to form a quasi-one-dimensional (1D) wire~\cite{thornton86} gives rise to quantization of conductance~\cite{wharam88,vanWees88} in units of \(2e^2/h\), which has been shown to be unaffected by the presence of weak electron-electron interactions.~\cite{maslov95} At low electron densities, long-range interactions dominate, resulting in a 1D Wigner crystal.~\cite{meyer08,schultz93} As the density increases, the Coulomb repulsion between electrons increases until it overcomes the confinement potential, whereupon the ground state distorts, which can lead to the bifurcation of the electronic system.~\cite{piacente04} The transition from a single- to a double-layer system as electron density is increased has previously been observed in two-dimensional (2D) electron systems confined to wide quantum wells,~\cite{shayegan} but the splitting of a 1D electron system into two rows has only recently been reported.~\cite{hew09}

We present transport measurements of weakly-confined quantum wires defined in a 2DEG by top-gated split-gate devices. In previous work, weakening the confinement potential led to the formation of two rows, marked by a jump in conductance $G$ from zero to \(4e^2/h\), although it was unclear whether there was coupling between the rows. Probing the transition into the double-row transport r\'egime, where a zigzag arrangement of electrons is expected,~\cite{klironomos07} we have now obtained clear evidence of coupling between the rows, showing anticrossing of bonding and antibonding states as the 1D confinement strength is tuned. Moreover, a different bias-induced structure is observed at 0.5\((2e^2/h)\) in addition to the usual one at 0.25\((2e^2/h)\), the second key result of this Rapid Communication. 

The conductance through two laterally aligned, but uncoupled, parallel wires formed by surface gates has been shown to be the sum of the conductance of each individual wire, resulting in plateaus at multiples of \(4e^2/h\).~\cite{smith89,simpson93} Vertically-aligned double quantum well (DQW) structures, where the accuracy of molecular-beam epitaxy growth allows very small inter-wire separation, have shown evidence of coupling between the parallel wires.~\cite{thomas99,castleton98} When there is strong coupling between wires, the electron wave functions hybridize, forming bonding and antibonding states, which manifest as anticrossings in the 1D subband energy levels. The minimum energy gap between the states occurs at the point of anticrossing and is given by \(\Delta_{\mathrm{SAS}}\), the energy difference between the symmetric and antisymmetric states. As the interlayer coupling is weakened, plateaus at multiples of \(4e^2/h\) begin to appear, where the energy levels of the two wells simply cross.~\cite{castleton98}

Our devices were fabricated using electron-beam lithography on 300 nm deep GaAs/AlGaAs heterostructures. Sample A consists of split gates, 0.4 \(\mu\)m long and 1 \(\mu\)m wide, and a top gate of width 1 \(\mu\)m defined above the split gates, separated by a 200 nm layer of cross-linked polymethylmethacrylate. After partial illumination, the carrier density and mobility were estimated to be \(1.5\times\!10^{11}\) cm\(^{-2}\) and \(1.3\times\!10^{6}\) cm\(^{2}\)V\(^{-1}\)s\(^{-1}\) respectively. Sample B has split gates 0.4 \(\mu\)m long and 1.9 \(\mu\)m wide, and a midline gate of width 1.1 \(\mu\)m in the plane of the split gates, with a \(0.4\) \(\mu\)m gap each side. After partial illumination, the carrier density and mobility were estimated to be \(1.9\times\!10^{11}\) cm\(^{-2}\) and \(3\times\!10^{6}\) cm\(^{2}\)V\(^{-1}\)s\(^{-1}\) respectively. The two-terminal conductance (\(G=dI/dV\)) was measured in a dilution refrigerator at 50 mK, with a 77 Hz excitation voltage of 5 \(\mu\)V. All data presented are from sample A unless otherwise stated.

\begin{figure*}[ht]
\begin{center}
\includegraphics[scale=0.43]{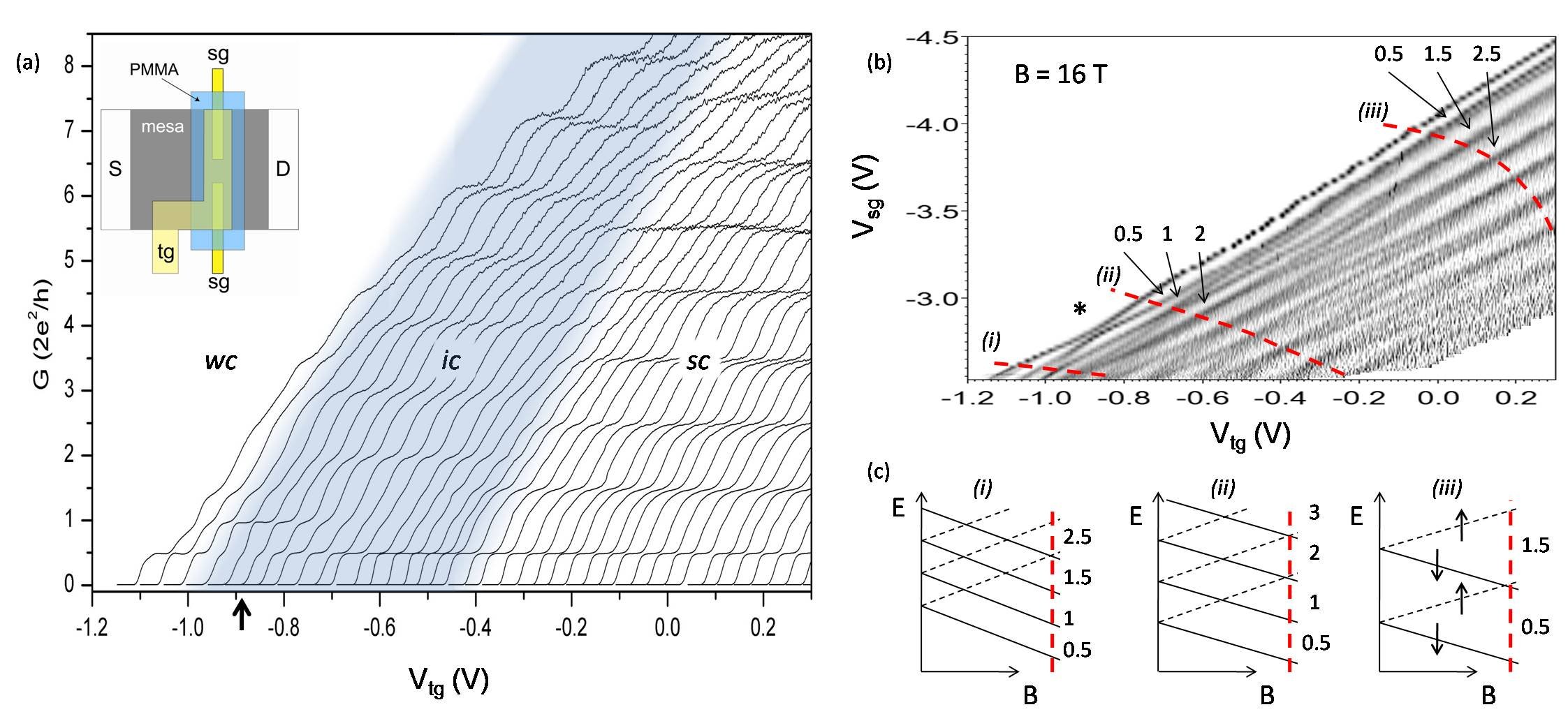}
\caption{(Color online) (a) Evolution of conductance plateaus with 1D confinement at $B$ = 16
T, the magnetic field was applied in the plane of the 2DEG, perpendicular to the 1D channel. Conductance \(G(V_{tg})\) was measured as \(V_{tg}\) was swept at fixed \(V_{sg}\), with a constant offset of 2 V between the split gates. \(V_{sg}\) was incremented between traces in steps of -50 mV (left to right) from -2.6,-0.6 V to -4.5,-2.5 V (voltages on both split gates given). The device layout is shown in the inset. The arrow indicates the region in which a double row forms. (b) Grayscale plot of transconductance \(dG/dV_{tg}\) as a
function of \(V_{tg}\) and \(V_{sg}\) at 16 T, with white representing regions of high transconductance. ``\(*\)'' marks the disappearance of the first plateau, and the anticrossing of energy levels. The dashed lines mark
typical ``cuts'' through each region of confinement; some of the plateau heights at these
cuts are marked. (c) Energy level diagrams showing the spin splitting of 1D subbands with magnetic field at points corresponding to (i), (ii), and (iii) on (b). Descending (solid) and ascending (dashed) branches represent spin-down ($\downarrow$) and spin-up ($\uparrow$) states, respectively, as illustrated in (c) (iii). The vertical dashed lines correspond to $B$ = 16 T.}
\label{figure1}
\end{center}
\end{figure*}

Split-gate devices with an added top or midline gate are versatile since the confinement strength and carrier density of the channel can be varied almost independently, allowing a number of new transport r\'egimes.~\cite{hew08} On sample A, control of the channel width was such that three successive conditions of spin-split energy level coincidence could be achieved at a fixed 16 T field.

Figure 1(a) shows conductance characteristics $G(V_{tg})$ for a range of fixed confinement strengths, determined by the split-gate voltage ($V_{sg}$). On the left of the figure, the 1D wire, just defined, is at its widest; and, by sweeping the top-gate voltage ($V_{tg}$) negatively, the carriers in the channel are depleted. Moving towards the right of Fig. 1(a), the width of the wire reduces as the 1D confinement strengthens, increasing energy gap between adjacent 1D subbands. For convenience, Fig. 1 has been divided into three confinement zones: strong (\emph{sc}), intermediate (\emph{ic}) and weak (\emph{wc}). At $B$=16 T, all 1D subbands are spin split, and the adjacent spin levels have crossed once (for example, in \emph{sc}) or several times, depending on the subband spacing. This is why, on moving from right to left through \emph{sc}, \emph{ic} and \emph{wc}, the quantization switches from odd to even and then back to odd integer multiples of $2e^2/h$, for $G>e^2/h$. This is a striking demonstration of the control achieved over the confinement potential. It was estimated using dc bias spectroscopy that the 1D subband spacing increased by a factor of 4 from weak to strong confinement.

\begin{figure*}[ht]
\begin{center}
\includegraphics[scale=0.35]{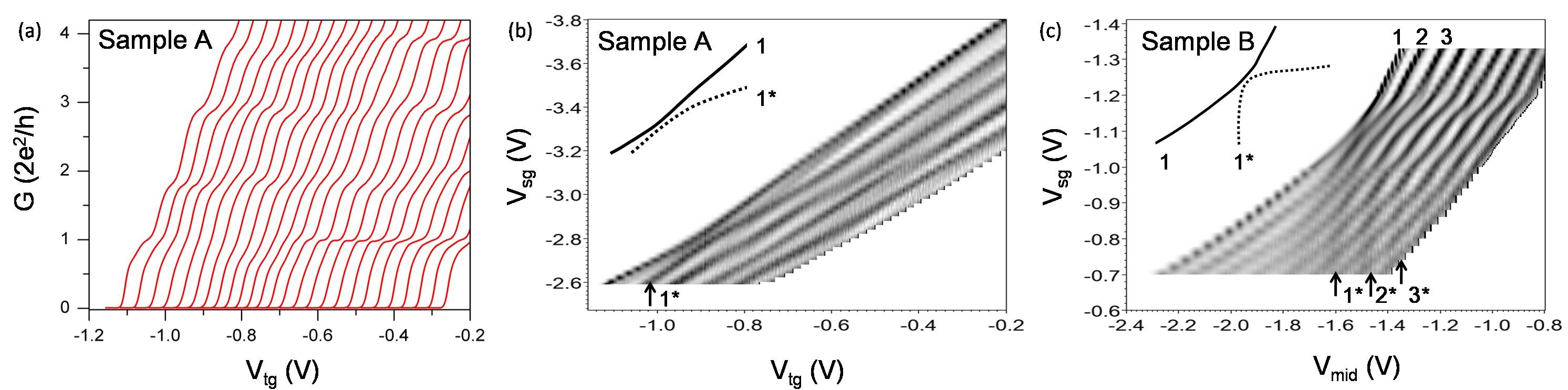}
\end{center}
\caption{(Color online) (a) Conductance traces at 0 T, sweeping $V_{tg}$ at fixed $V_{sg}$ (sample A). $V_{sg}$ is incremented from -2.6,-0.6 V (left) to -3.7,-1.7 V (right). (b) Grayscale \(dG/dV_{tg}\) plot of (a) as a function of  $V_{tg}$ and $V_{sg}$. The first bonding (1) and antibonding (1*) states are shown in the inset. (c) Grayscale \(dG/dV_{mid}\) plot for sample B as a function of $V_{mid}$ and $V_{sg}$.  The first bonding (1) and antibonding (1*) states are shown in the inset; the first three bonding
and antibonding states are marked on the grayscale.}
\label{figure2}
\end{figure*}

The grayscale in Fig. 1(b), plotting the transconductance ($dG/dV_{tg}$) against $V_{sg}$ and $V_{tg}$, shows how the subband energy levels cross as the confinement strength changes. The white regions correspond to plateaus and black lines to the risers between them. Dashed lines (i), (ii), and (iii) are typical ``cuts'' in the $V_{sg}-V_{tg}$ plane for the three r\'egimes. These are described by the schematic diagrams in Fig. 1(c), which show how the energy levels split and cross with magnetic field through those ``cuts''. The numbers indicate the plateaus that would be observed were the energy levels to be populated along the vertical dashed lines.

In a magnetic field we would expect the plateau at $e^2/h$ to remain strong regardless of confinement strength. However, we observe a weakening of this plateau at \(V_{tg}\) \(\approx\) -0.9 V. In a previous study, we reported a complete disappearance of the first plateau, indicative of the formation of a double row.~\cite{hew09} Here, we show that the weakened first plateau reflects coupling between the rows, and the rest of this Rapid Communication is devoted to characterizing the coupling behavior further. The split-gate width of sample A is 1 $\mu$m, compared to 0.7 $\mu$m previously;~\cite{hew09} thus it has been possible to more finely tune the carrier densities and channel widths in the transition region between one and two rows. The weakened first plateau corresponds to the anticrossing of the first and second spin-down subbands, marked by the asterisk in Fig.~1(b). This is attributable to the hybridization of the wave functions, which form bonding and antibonding states. The anticrossing behavior has not previously been observed in a single quantum wire since it depends crucially on the interaction and spatial distribution of the electrons.

In \emph{ic}, the $2e^2/h$ plateau weakens from $V_{tg}>-0.8$~V. This is reflected in Fig. 1(b), where the 1$\uparrow$ and 2$\downarrow$ levels cross over a larger range of $V_{sg}$ than higher subband crossings, perhaps a consequence of row formation. 

Figure 2 details how the conductance and transconductance evolve with confinement strength at $B = 0$~T. Figures 2(a) and 2(b) correspond to the top-gated device (sample A) and Fig. 2(c) to the midline-gated device (sample B). Figure 2(a) shows a weakening of the first conductance plateau $(2e^2/h)$ with weakening confinement, accompanied by a faint structure around $0.7(2e^2/h)$. A plateau appears at $2e^2/h$ at the weakest confinement (on the left), reflecting a return to a single row of electrons. This is a manifestation of the interplay between weakening confinement and lowering density.

The weakening of the first quantized plateau in zero and finite magnetic fields is unexpected since the standard subband model provides no mechanism for it: the energy of the first excited state must be greater than that of the ground state. As the carrier concentration decreases and confinement weakens, the energy gap between the these two states shrinks to near-degeneracy, whereupon they hybridize into bonding and antibonding states. We suggest that the ground-state wave function has been progressively distorted by the increasing strength of interaction. The correlated motion of electrons may produce a zigzag configuration,~\cite{klironomos07} eventually separating into two parallel conducting rows.

The grayscale of Figs. 2(b) and 2(c) show a modulation of the dark lines in the region of coupling between rows -- the formation of bonding and antibonding states in coupled wire systems~\cite{thomas99,salis} is manifested in this way -- such that it is possible to discern the antibonding states as superimposed parabolae. The inset of Fig. 2(c) schematically represents the first bonding-antibonding pair of states, and arrows on the grayscale mark the first three antibonding states clearly seen in sample B. Only one such state is clear in sample A [Fig. 2(b), inset].

The coupling strength between two quantum wires depends on the overlap of their respective electron wave functions. In vertically coupled wires,~\cite{castleton98,thomas99} the center-to-center separation of the two quantum wells in the $z$ direction was of order 20~nm. It is difficult to reproduce such closely spaced parallel wires in a single 2DEG by electrostatic gating due to limitations imposed by lithographic resolution and the distance between the 2DEG and the gates. In our results, the two rows are produced by interactions within the channel: changing the carrier density and confinement strength thus tunes the coupling. Before the electrons divide into two rows, a zigzag arrangement is expected, where theory predicts that a number of possible phases can exist.~\cite{klironomos07} 

The energy gap between the symmetric and antisymmetric states was estimated to be \(\Delta_{\mathrm{SAS}}\sim\!0.2\)~meV in sample A. The anticrossing in Fig.~\ref{figure1}(b) shows that the two rows remain coupled even in high fields. In DQW systems, high in-plane perpendicular fields were shown to completely decouple the wires,~\cite{thomas99} the mechanism for which was the shifting of the Fermi circles in each 2DEG with respect to the other. However, our system has a single 2DEG with the rows coupled laterally. A crucial difference between these two studies is that the electrons are tightly confined to their respective wells by the band structure in the DQW, whereas, in our case, the electron rows are weakly confined in the lateral direction, separated by their own weak Coulomb barrier. Thus the bonding and antibonding states may correspond to the transverse modes that can be excited in the two rows.~\cite{meyer07}

\begin{figure}[t]
\begin{center}
\includegraphics[scale=0.25]{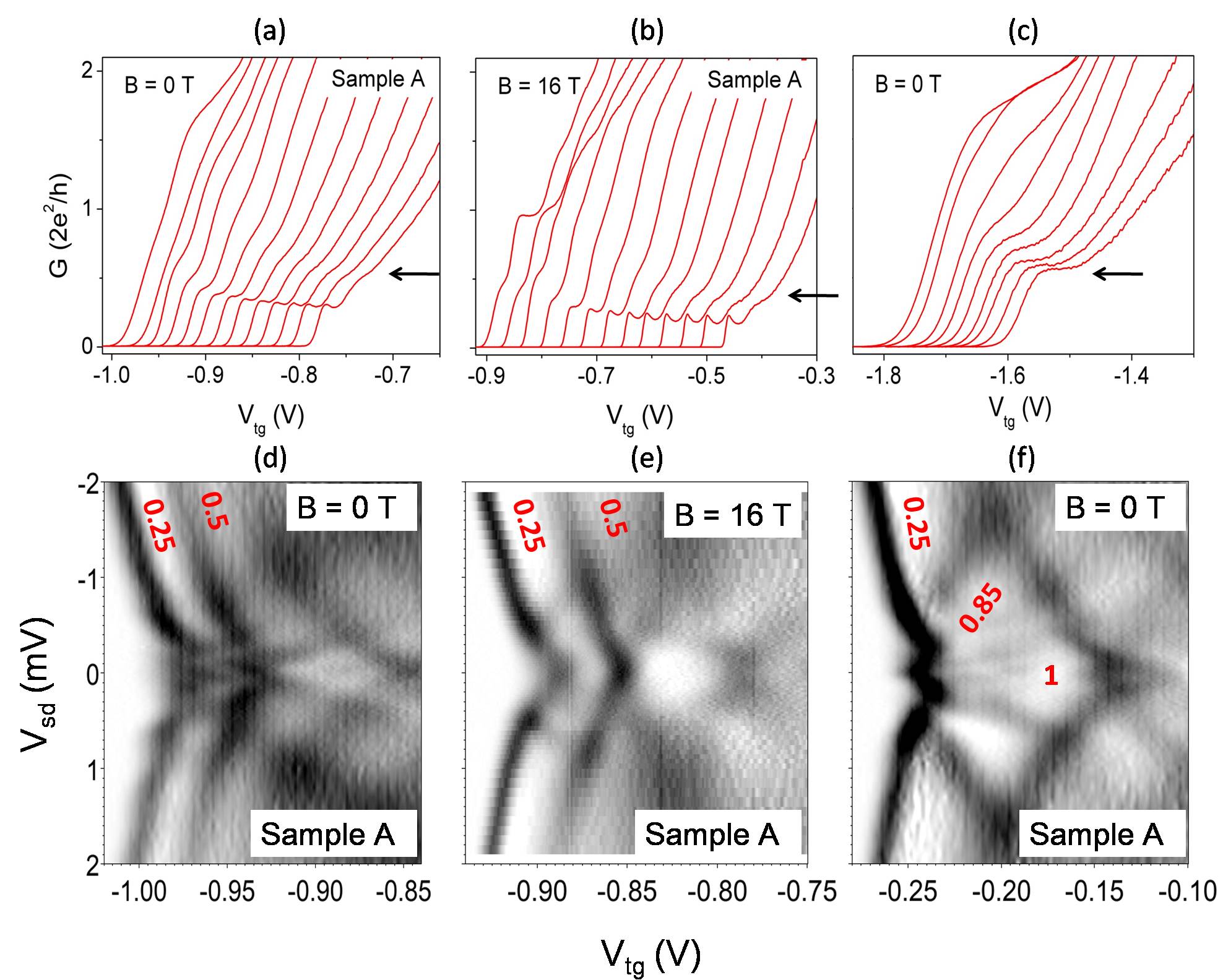}
\end{center}
\caption{(Color online) (a) and (b) Conductance traces sweeping \(V_{tg}\) at $V_{sg}$ =
-2.59,-0.77 V ($B = 0$ T) and -2.79,-0.79 V ($B = 16$ T), respectively.
\(V_{sd}\) is incremented from zero on the left to -3 mV on the right, in steps
of 0.25 mV; traces are offset for clarity. (c) Conductance traces at $B =
0$ T, sweeping $V_{sg}$, for an uncoupled double row of electrons (data from another quantum wire).
$V_{sd}$ is incremented from left to right in steps of 0.5 mV from zero to -5 mV;
traces are offset for clarity. The arrows indicate the 0.5 feature at finite bias.
The grayscale plots of \(dG/dV_{tg}\) corresponding to (a) and (b) are shown in (d) and (e), respectively, as a function of \(V_{tg}\) and $V_{sd}$. Grayscale (f) shows the
characteristic single row behavior, in strong confinement, at 0 T. Certain conductance plateaus are marked on the grayscales in units of \(2e^2/h\).}
\label{figure3}
\end{figure}

In general, conductance features in a finite dc source-drain bias are not well understood for $G <2e^2/h$. A strong structure at $0.25(2e^2/h)$ is a unique feature of a single quantum wire under high bias.~\cite{chen08} In our sample, we see two structures when the rows are coupled, one at 0.25 and the other, rather weaker structure, at 0.5. Figure 3(a) shows the transition that occurs in conductance traces for the coupled rows, from zero to finite source-drain bias voltage ($V_{sd}$). The same behavior is observed in Fig.~3(b) at 16 T. Figure 3(c) shows a strong lone structure at 0.5 with high $V_{sd}$, measured in a similar quantum wire, but where the two rows were uncoupled. This 0.5 feature is the simple addition of the 0.25 from two independent rows and remains at 0.5 in high magnetic fields. We may therefore infer that the two structures in (a) and (b) are a result of the coupling of the rows. Unlike the linear regime [Figs. 1(a) and 2(a)] in which the plateau marking the bonding state is weaker, at high $V_{sd}$ the bonding state (marked by the 0.25 feature) seems stronger than the antibonding state (0.5 feature). A stronger 0.25 feature implies an increased coupling between the two rows, reflecting a tendency towards single row transport with increasing source-drain bias. We speculate that this may occur through the intermediary of a zigzag arrangement of electrons.

Figures 3(d) and (e) are grayscale diagrams corresponding to (a) and (b), with the 0.25 and 0.5 structures labeled. There is no difference in the occurrence of the structures in zero or finite magnetic field, as is also the case for a single row. Figure 3(f) corresponds to the same sample but in the \emph{sc} r\'egime where single wire behavior is observed, showing the usual 0.25 feature.

In conclusion, we have shown direct evidence of the formation of an interacting double-row system in a quasi-1D channel at weak confinement and low electron density. Coupling between the rows was marked by the anticrossing of energy levels at both $B$=0 and 16 T. Bonding and antibonding states of the coupled rows were observed in two different devices. The structure seen at $0.25$ and $0.5(2e^2/h)$ in high dc bias shows that the coupling persists and indeed strengthens at high bias. With the weakening of the first plateau at low confinement strengths, we enter the r\'egime wherein a zigzag structure is predicted. The anticrossing may therefore be an indirect signature of such a spatial configuration.

This work was supported by the Engineering and Physical Sciences Research Council. W.K.H. acknowledges support from Toshiba.

\end{document}